# Pulsed Electron Holography

Matthias Germann, Tatiana Latychevskaia*, Conrad Escher, Hans-Werner Fink

Physics Institute, University of Zurich

Winterthurerstrasse 190

CH-8057 Zurich

Switzerland

**ABSTRACT**

A technique of pulsed low-energy electron holography is introduced that allows for recording highly resolved holograms within reduced exposure times. Therefore, stacks of holograms are accumulated in a pulsed mode with individual acquisition times as short as 50μs. Subsequently, these holograms are aligned and finally superimposed. The resulting holographic record reveals previously latent high-order interference fringes and thereby pushing interference resolution into the sub-nanometer regime. In view of the non-damaging character of low-energy electrons, the method is of particular interest for structural analysis of fragile biomolecules.

*Corresponding author: Tatiana Latychevskaia, e-mail: tatiana@physik.uzh.ch

Holography, among other imaging techniques, comprises a solution to the phase problem and allows a complete three-dimensional restoration of the recorded objects in just one reconstruction step[1]. Robust, sub-nanometer-sized objects are routinely imaged nowadays by means of high-energy electron holography[2-3]. In addition to such high-energy electron microscopy, holography with low energy electrons in the 100 eV range has been realized in a lens-less setup[4] and applied to fragile objects like individual DNA molecules[5]. It has been shown that even a vast dose of $10^8$ electrons/nm$^2$ leaves DNA molecules unperturbed[6]. This demonstrates that coherent low-energy electrons are the only non-damaging Ångstrom wavelengths radiation. Although, lenses and their intrinsic aberrations are avoided in low-energy electron holography, the experimental resolution is currently limited to just about 1 nm due to residual vibrations, ac-magnetic fields and other experimental instabilities[7]. These experimental disturbances smear out the highest order interference fringes[8] which however, carry the fine details of the molecule's structure. In view of structural analysis, it is thus of great importance to record highly resolved interference patterns. Accordingly, the sharpness of a hologram can be improved by decreasing the acquisition time. But this in turn leads to a poor signal-to-noise ratio (SNR). Here we present a way out of this conflict - pulsed electron holography combined with image registration algorithms.

A low-energy electron point source (LEEPS) microscope designed for inline holography[5-6] has been modified to operate in the pulsed mode. In the LEEPS microscope, sketched in Fig. 1(a), a sharply pointed field emitter provides a source of coherent low-energy electrons. The sample positioned into the divergent electron beam at a distance of about 1 μm from the source scatters part of the wave to form the hologram by interference with the un-scattered part of the wave. The hologram is captured by a detector unit placed at a distance of about 10 cm from the source. In the conventional mode the sample is continuously exposed to electron radiation while the holograms are acquired continuously with standard video frame rate using a CCD camera.

For the acquisition in the pulsed mode a sequence of three temporally overlapped pulses, as drafted in Fig. 1(b), allows for both, short acquisition time as well as reduced exposure time. The CCD camera monitors the detector screen for 20 ms; within that time interval the field emitter is set on high potential for 2 ms. Hence the sample is exposed to electrons only for that period of time. Again, in the interval of these 2 ms the detector unit is switched on for an actual acquisition time of just 50 µs. Such succession of time intervals not only affords short acquisition times but also preserves the coherence of the imaging electrons as constant tip voltage and associated deBroglie wavelength is established well before the hologram is actually acquired.

In this manner, individual holograms are recorded repeatedly and accumulated to stacks of typically a few hundred records. The interval between successive acquisitions is of the order of one second necessary to read out the CCD camera memory.

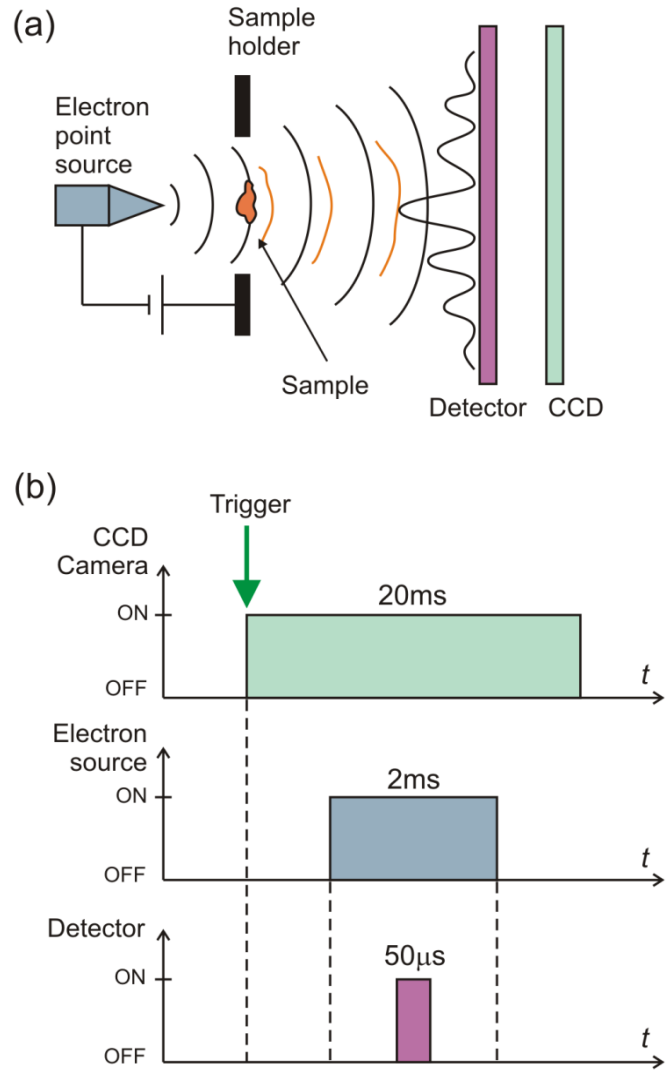

FIG. 1. Principle of pulsed low-energy electron holography. (a) Schematic of the microscope. The freestanding sample is exposed to a coherent electron wave and the interference pattern formed in the detector plane is monitored with a CCD camera. (b) Sketch of the temporal sequence of voltage pulses used for the record of a single individually pulsed hologram (IPH). Within 20 ms monitor time the sample is exposed to electrons for 2 ms and the holographic data are acquired for 50 μs. Note that the pulses are not drawn to scale.

To demonstrate pulsed electron holography, holograms of λ-DNA molecules are recorded. The samples are prepared in compliance with the procedure of plunge freeze-drying known from transmission electron microscopy. First, a thin film of dilute DNA solution is applied onto a hydrophilized holey carbon membrane. Next, the film is vitrified by plunging it into liquid ethane and finally freeze-dried. After the transfer into the LEEPS microscope, the membrane is scanned to search for free standing DNA-molecules.

As described earlier, data acquisition in the pulsed mode generates stacks of individually pulsed holograms (IPH). These holograms are then aligned by image registration algorithms[9]. To compensate for mechanical drift we correct the relative shift between holograms using a method based on finding the location of the cross-correlation maximum[10]. In order to avoid the distracting influence of screen defects or detector edges the cross-correlation procedure is only applied to a selected region free from distorted image data. The aligned holograms are superimposed eventually. The correspondent compiling loop includes the following steps:

(i) An appropriate region $\rho_1$ is defined within the first IPH (IPH1, reference hologram) of the stack.

(ii) The second IPH (i.e. IPH2) is read from the stack and the corresponding region $\rho_2$ is assigned.

(iii) The cross-correlation is performed between $\rho_1$ and $\rho_2$ and the position of the cross-correlation maximum determines the misalignment correction for IPH2 relative to IPH1.

(iv) The *entire* IPH2 is shifted accordingly and superimposed onto IPH1.

Compiled like that, IPH1 and IPH2 provide the next reference hologram for the routine to continue along step (i)-(iv) and successively utilize the data of all subsequent IPHs. In addition, low-pass filtering is applied to the selected regions $\rho_i$ between the steps (ii) and (iii) to suppress statistical noise and thus improve the precision in determining the coordinates of the cross-correlation maximum.

The normalized two-dimensional cross-correlation function between two holograms of MxN pixels in size is calculated as[11]:

$$I_{IPH1} \otimes I_{IPH2}(n,m) = \frac{\sum_{i,j}^{N,M}(I_{IPH1}(i-n,j-m) - \bar{I}_{IPH1})(I_{IPH2}(i,j) - \bar{I}_{IPH2})}{\sqrt{\sum_{i,j}(I_{IPH1}(i,j) - \bar{I}_{IPH1})^2}\sqrt{\sum_{i,j}(I_{IPH2}(i,j) - \bar{I}_{IPH2})^2}} \qquad (1)$$

where $I_{IPH}(i,j)$ is the intensity of pixel (i,j) of the IPH, and $\bar{I}_{IPH}$ is the mean intensity of the IPH: $\bar{I}_{IPH} = \frac{1}{MN}\sum_{i,j}^{N,M} I_{IPH}(i,j)$.

The numerator of Eq.(1) is calculated with the help of a Fourier-transform:

$$I_{IPH1} \otimes I_{IPH2} = F^{-1}\left(F^*(I_{IPH1} - \bar{I}_{IPH1})F(I_{IPH2} - \bar{I}_{IPH2})\right) \qquad (2)$$

Since the signal $I_{IPHk}$ corresponds to a real number, the relation $F^*(I_{IPHk}) = F^{-1}(I_{IPHk})$ is applicable. An interpolation of the cross-correlation distribution improves the precision of locating its maximum[12].

First, we compare a conventionally recorded hologram with a single corresponding IPH and with the compilation of a stack of such IPHs, respectively. All holograms are recorded with electrons of 168 eV kinetic energy. Figure 2 shows (a) the conventionally recorded hologram of DNA molecules with 20ms acquisition time, (b) an IPH of the same sample with 50 μs acquisition time and finally (c) the compilation of 74 aligned IHPs with an effective acquisition time of 3.7 ms in total. As expected, the single IPH is much noisier than the conventionally recorded hologram considering the different acquisition times. However, as apparent from visual inspection of the zoomed-in areas in Fig. 2(a) and (c) the compilation of all IPHs clearly features more signal than the conventionally recorded hologram, despite the reduced acquisition time. The shifts Δx and Δy of the IPHs range from 0 to 45 pixels with respect to the corresponding reference hologram. In Fig. 2(d) the misalignment correction $\Delta r = \sqrt{\Delta x^2 + \Delta y^2}$ of all the IPHs in the stack is plotted in ascending order.

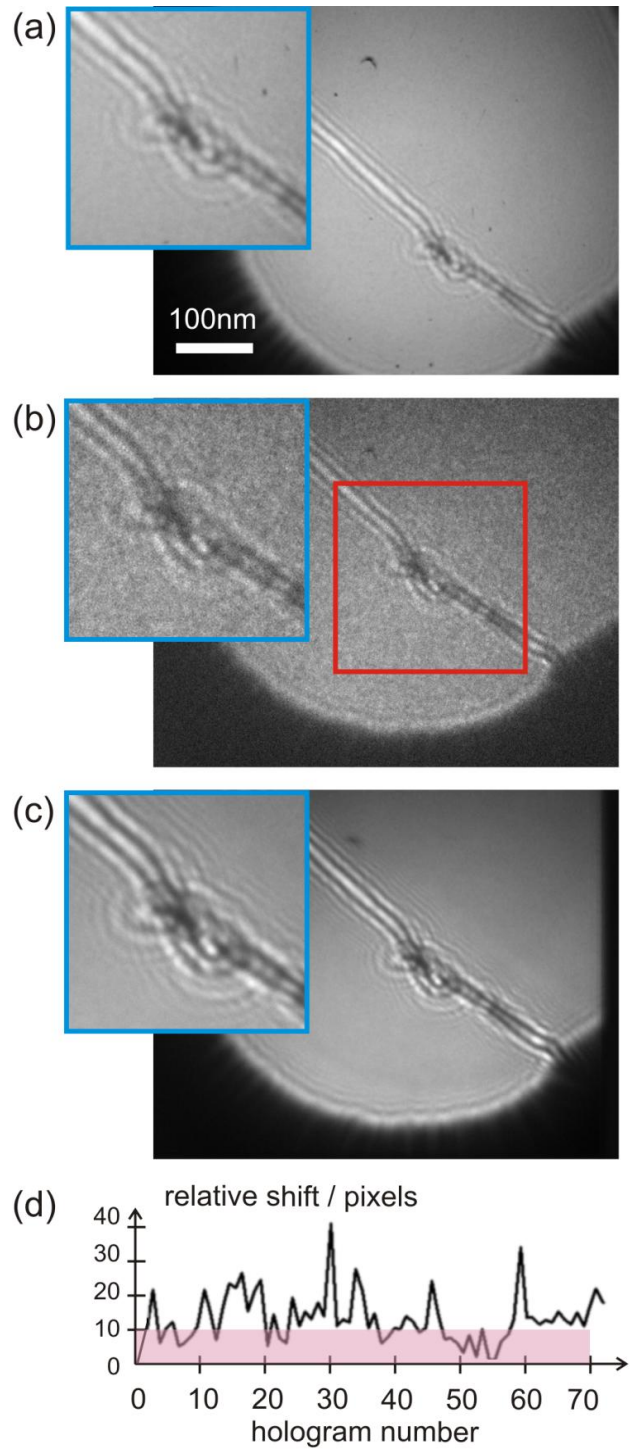

FIG. 2. Holograms of DNA molecules acquired in different modes using 168 eV electrons. The inlets (blue frames) show the magnified central area of the holograms. (a) Conventionally recorded hologram with an acquisition time of 20 ms. (b) One IPH acquired in the pulsed mode within 50 µs. The area marked in red is used for the cross-correlation calculation and the subsequent superposition of several IPHs. (c) Compilation of a stack of 74 IPHs. (d) Misalignment correction (in pixels) of each IPH relative to the 1st hologram. The pink stripe displays the limit of the spacing between the finest fringes in the hologram shown in Fig. 2(c) and corresponds to 10 pixels.

For a quantitative evaluation of the improvement of resolution, a stack of 500 IPHs of DNA molecules is recorded with 140 eV electrons. An individual IPH is shown in Fig. 3(a) with the corresponding intensity profile in Fig. 3(c), demonstrating that only lower interference orders up to second order at most can be distinguished from noise. Figure 3(b) shows the compilation of 500 IPHs. The intensity profile for the compilation, displayed in Fig. 3(d), shows that up to at least 7$^{th}$ order interference fringes can be recognized and that the SNR has significantly improved.

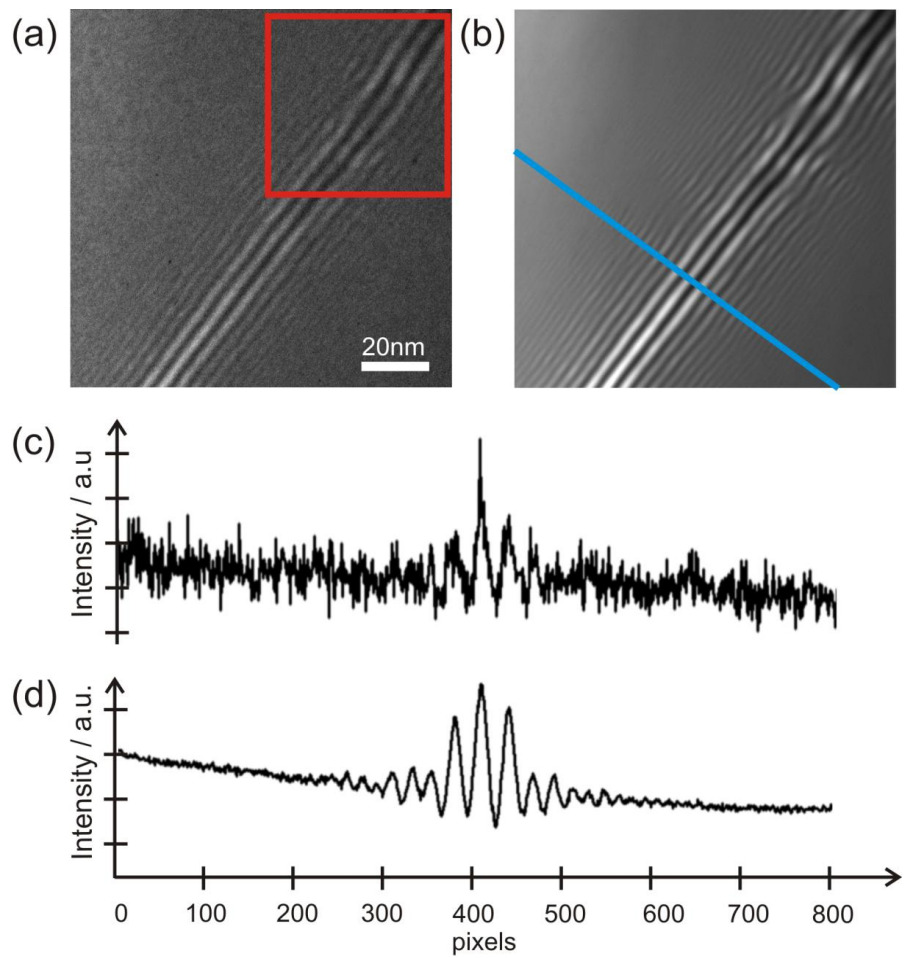

FIG. 3. Quantitative evaluation of the spatial resolution. (a,b) Single IPH of DNA molecules and the compilation of a stack of 500 IPHs, respectively. The blue line indicates the direction of the line cuts shown in (c) for the IPH and in (d) for the stack, accordingly. The SNR in (c) is obviously worse than the one in (d). The separation of the highest order interference fringes identifiable in the intensity profiles can be used as a criterion for resolution and lead to 12 Å for the IPH and an improved resolution of 6 Å for the compiled stack. The area captured in the red squares shows the region selected for the cross-correlation calculation.

In analogy to X-ray crystallography, where the resolution is defined by the distance between the zero-order spot and the highest visible diffraction peak, in holography, the resolution is defined by the distance between the zero-order fringe and the finest detectable high-order interference fringe. The hologram of a point-scatterer is a Fresnel zone pattern, and its reconstruction is described by a δ-function if an ideal infinite screen is assumed to record the hologram. Experimentally, the finite size $S$ of the detector leads to a reconstruction described by a sinc- instead of a δ-function. For a hologram of two closely placed point sources, its reconstruction shows two overlapping sinc-functions and thus, the Rayleigh criterion of resolution can be applied. The distance between the zero-order maximum and the first minimum of the sinc-function constitutes the resolution limit and equals to $R= \lambda z/S$, where $z$ is the distance between object and screen and $S$ amounts to twice the distance between the zero-order fringe and the highest order fringe identifiable in the hologram. This fairly general interference resolution criterion has also been applied to the theory of low-energy electron holography and verified by cluster calculations of holograms and their reconstructions[13].

Typical values for our setup are: $\lambda$=0.1 nm, a sample screen distance $z$ of 10 cm and a distance $S$=10 mm, result in a resolution of approximately 1 nm.

In fact, the IPH in Fig. 3(a) exhibits a resolution of 12 Å and the compilation displayed in Fig. 3(b) reaches a resolution of 6 Å. Further improvement in the alignment procedure might be accomplished by applying subpixel registration methods[12].

As illustrated above, the enhancement of resolution is closely related to an increase of the SNR. The latter shall thus be analyzed in more detail, also to get an idea about the reasonable stack-size in view of a sufficient SNR. The SNR for an entire hologram is not directly available due to the strong variations in intensity within a holographic record. Instead, the SNR is evaluated for a selected region, where - apart from statistical and experimental fluctuations - the intensity is expected to be constant and of a certain average value $I_0$ (see square region marked in yellow in Fig. 4(a)). The pixel intensity fluctuations within this region are plotted in form of a histogram shown in Fig. 4(b). Evidently, the statistical noise exhibits a Gaussian distribution. From the accordant fit with a Gaussian function the parameters $\bar{I}$ (mean value) and $\sigma$ (standard deviation) can be extracted. The SNR is then given by $\mathrm{SNR} = \bar{I}/\sigma$. In the case of the single IPH depicted in Fig. 4(a), this leads to the numerical values: $\bar{I}_{\mathrm{IPH}} = 36.71$, $\sigma_{\mathrm{IPH}} = 11.95$, and thus to $\mathrm{SNR}_{\mathrm{IPH}} = 3.07$. Throughout the entire stack of 500 IPHs the $\mathrm{SNR}_{\mathrm{IPH}}$ varies between 1 and 3.5.

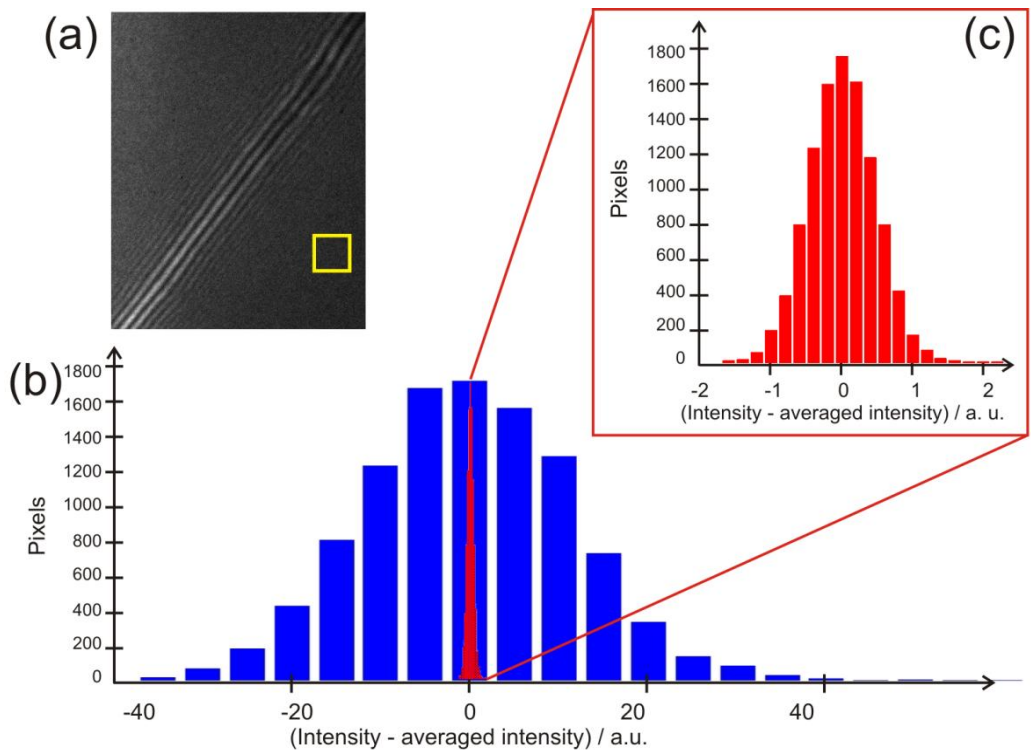

FIG. 4. Evaluation of noise. (a) In the IPH of DNA molecules, the area marked in yellow (100x100 pixels) is selected for analyzing the noise distribution at constant background intensity. (b) Histogram of the pixel intensity fluctuations within the area in question in an IPH. The data are Gaussian distributed. (c) The histogram plotted in red shows the pixel intensity distribution within the corresponding area of the 500 compiled IPHs. Again, the data exhibit a Gaussian distribution but with a much smaller deviation indicating an increase in the SNR.

The pixel intensity fluctuation distribution within the corresponding region of the 500 compiled IPHs is shown in Fig. 4(c). The numerical analysis reveals a mean value of $\bar{I} = 15.70$ and a standard deviation of $\sigma = 0.47$. The SNR thus amounts to 33.42.

Within a selected region of average intensity $I_0$, as discussed above, the mean intensities $\bar{I}_{IPHk}, k = 1...P$ and the standard deviations $\sigma_{IPHk}$ for all IPHs may vary throughout a stack of stack-size P. In the course of compiling IPHs a sequence of stacks with increasing stack-size P is generated. For a stack of P compiled IPHs, the intensity of pixel (i,j) amounts to:

$$I^{(P)}(i,j) = \frac{1}{P} \sum_{k=1}^{P} I_{IPHk}(i,j) \quad (3)$$

The mean intensity and the standard deviation are given by the expressions $P\bar{I}^{(P)} = \bar{I}_{IPH1} + \bar{I}_{IPH2} + ... + \bar{I}_{IPHP} \approx P\bar{I}_{IPHn}$ and $P^2 \left(\sigma^{(P)}\right)^2 = \sigma_{IPH1}^2 + \sigma_{IPH2}^2 + ... + \sigma_{IPHP}^2 \approx P\sigma_{IPHn}^2$ correspondingly, where IPHn denotes any representative IPH from the stack.

To these stacks the Central Limit Theorem applies and thus, the pixel intensities $I^{(P)}(i,j)$ are Gaussian distributed around $\bar{I}^{(P)}$ with the standard deviation $\sigma^{(P)}$, where $\bar{I}^{(P)} \to I_0$ with $P \to \infty$.

Hence, the SNR increases proportional to $\sqrt{P}$:

$$SNR^{(P)} = \frac{\bar{I}^{(P)}}{\sigma^{(P)}} = \frac{\bar{I}_{IPHn} \sqrt{P}}{\sigma_{IPHn}} = SNR_{IPHn} \sqrt{P} \quad (4)$$

Eq. (4) can thus be used to easily estimate how many holograms with $SNR_{IPHn}$ must be compiled to achieve a desired SNR. The SNR dependence for a series of 500 IPHs of DNA is shown in Fig. 5. While 500 IPHs are being compiled, the SNR increases from 3.07 to 33.42 and follows a square-root dependence on the number of IPHs. The slight fluctuations in the SNR curve are due to intensity variations within subsequent IPHs.

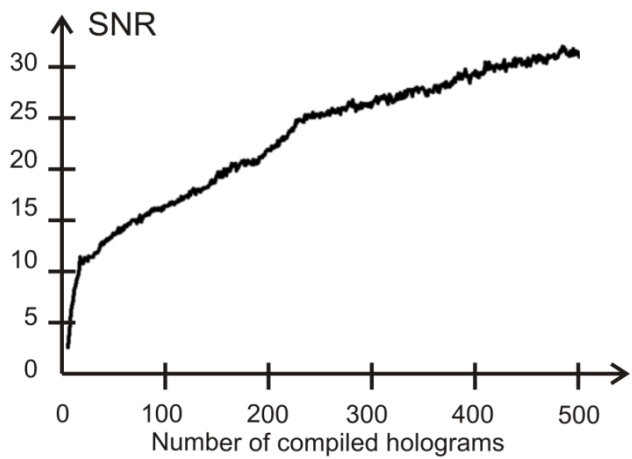

FIG. 5. Evolution of the SNR with increasing stack-size. The graph shows the evolution of the SNR in the course of the iterative procedure of compiling a total of 500 IPHs. The SNR is evaluated for the 100x100 pixel area marked in yellow in Fig 4(a). The data reflect the increase pursuant to $\sqrt{P}$, where P is the number of successively compiled IPHs.

Driven in the conventional mode, LEEPS holography is reliant on very steady experimental conditions. Acquiring IPHs in the pulsed mode now affords to reduce the actual acquisition time by three orders of magnitude and thus to exclude perturbation due to mechanical instabilities. In order to compensate for the poor SNR of one single IPH, hundreds of IPHs can be compiled. Hence, the effective acquisition time is accumulated accordingly. The result is a holographic record with an optimized resolution compared to conventional electron holograms. In fact, the spatial resolution estimated from the distance between the highest order interference fringes amounts to about 6 Å. The quantitative analysis of the SNR shows a square-root dependence on the number of compiled IPHs. Apart from improving interference resolution in holograms, the technique of pulsed holography also offers the possibility to carry out pump-probe experiments[14] with high spatial and temporal resolution to address such phenomena like charge hopping processes in biological molecules.

**ACKNOWLEDGEMENTS**

We gratefully acknowledge financial support of the Swiss National Science Foundation.We gratefully acknowledge financial support of the Swiss National Science Foundation.